\documentclass[aps,prd,superscriptaddress,nofootinbib,showkeys,twocolumn,preprintnumbers]{revtex4-1}
\pdfoutput=1

\usepackage[utf8]{inputenc}
\usepackage{uniinput}
\usepackage{graphicx}
\usepackage{color}
\usepackage{amsmath,amssymb,amsthm,mathtools,dsfont,amsfonts}
\usepackage{comment,braket,array,bold-extra,cancel,soul}
\usepackage{enumerate,xargs}
\usepackage{hyperref}
\usepackage[plain]{fancyref}
\usepackage{subcaption}
\usepackage[dvipsnames]{xcolor}
\usepackage{fourier-orns}

\newcommand{\D}{\mathrm{d}}

\newcommand{\bea}{\begin{eqnarray}}
\newcommand{\eea}{\end{eqnarray}}
\newcommand{\be}{\begin{equation}}
\newcommand{\ee}{\end{equation}}

\usepackage[format=plain,
textfont=it
]{caption}
\usepackage{microtype}

\usepackage{graphicx}
\graphicspath{{figures/}}

\begin{document}
\widetext
\leftline{}
\centerline{\em This essay received an Honorable Mention in the 2021 Essay Competition of the Gravity Research Foundation.}
\noindent\makebox[\linewidth]{\rule{\textwidth}{0.5pt}} 

\title{Curing with hemlock: escaping the swampland using instabilities from string theory}

\author{Souvik Banerjee}
\email{souvik.banerjee@physik.uni-wuerzburg.de}
\affiliation{Institut für Theoretische Physik und Astrophysik,
	Julius-Maximilians-Universität Würzburg, Am Hubland, D-97074 Würzburg, Germany}

\author{Ulf Danielsson}
\email{ulf.danielsson@physics.uu.se}
\affiliation{Institutionen för fysik och astronomi, Uppsala Universitet, Box 803, SE-75108 Uppsala}

\author{Suvendu Giri}
\email{suvendu.giri@unimib.it}
\affiliation{Dipartimento di Fisica, Universit'a di Milano-Bicocca, I-20126 Milano}
\affiliation{INFN, sezione di Milano-Bicocca, I-20126 Milano, Italy}

\begin{abstract}
	\noindent
	In this essay we will take a wonderful ride on a dark bubble with strings attached, which carries our universe out of the swampland and makes it realizable in the landscape of string theory. To find the way to the landscape, we make use of apparently hostile corners of the swampland and their instabilities.

	\begin{center} 
		\rule{10pt}{0.3pt} {\fontencoding{U}\fontfamily{futs}\selectfont\aldine\relax}
		 \rule{10pt}{0.3pt} 
	\end{center}
\end{abstract}

\preprint{UUITP-15/21}

\maketitle

\section*{Introduction}

The discovery at the turn of the millennium, that we live in an accelerating universe was a revolution in cosmology and fundamental physics. Current observations indicate that the cause is a ``dark energy'' with an equation of state consistent with Einstein's cosmological constant.

String theory is our best bet for a fundamental UV complete theory, and the struggle to obtain a positive cosmological constant (de Sitter universe: dS) in string theory began soon after the observational discovery. Unfortunately, compactifications of string theory to lower dimensions naturally lead to a negative value of the cosmological constant (anti de Sitter universe: AdS). Despite encouraging results in the early 2000s, finding reliable dS solutions in string theory remains an open problem, and the search for even a single fully consistent example of a vacuum with a positive cosmological constant is the subject of many ongoing efforts. See \cite{Danielsson:2018ztv}, and references therein, for an overview.

The absence of any reliable clues from string theory on where to look for such a dS vacuum has led to the ``swampland program'' with the goal of finding out \emph{where not to look}. This program attempts to find key properties that distinguish between effective theories that can be obtained from compactificatied string theory or quantum gravity in general (said to be in the \emph{landscape}), and those that cannot (said to lie in the \emph{swampland}). See \cite{Palti:2019pca} for an indepth review. 

One of the conjectures that has been put forward by this program is that there are \emph{no dS vacua in string theory} -- not even metastable ones. AdS vacua, on the other hand, are common in string theory but there are reasons to believe that  unless they are supersymmetric,  they are unstable and must decay either perturbatively on non-pertubatively, \cite{Ooguri:2016pdq, Freivogel:2016qwc}. 

Dark energy poses a challenge for string theory, and can be viewed as its first observational test. The question to ask is -- \emph{Is there a way out of this swamp?} 
\begin{figure*}
	\centering
	\begin{subfigure}[b]{0.45\textwidth}
		\centering
		\def\svgwidth{0.7\linewidth}
\begingroup%
  \makeatletter%
  \providecommand\color[2][]{%
    \errmessage{(Inkscape) Color is used for the text in Inkscape, but the package 'color.sty' is not loaded}%
    \renewcommand\color[2][]{}%
  }%
  \providecommand\transparent[1]{%
    \errmessage{(Inkscape) Transparency is used (non-zero) for the text in Inkscape, but the package 'transparent.sty' is not loaded}%
    \renewcommand\transparent[1]{}%
  }%
  \providecommand\rotatebox[2]{#2}%
  \newcommand*\fsize{\dimexpr\f@size pt\relax}%
  \newcommand*\lineheight[1]{\fontsize{\fsize}{#1\fsize}\selectfont}%
  \ifx\svgwidth\undefined%
    \setlength{\unitlength}{278.77707258bp}%
    \ifx\svgscale\undefined%
      \relax%
    \else%
      \setlength{\unitlength}{\unitlength * \real{\svgscale}}%
    \fi%
  \else%
    \setlength{\unitlength}{\svgwidth}%
  \fi%
  \global\let\svgwidth\undefined%
  \global\let\svgscale\undefined%
  \makeatother%
  \begin{picture}(1,0.98226738)%
    \lineheight{1}%
    \setlength\tabcolsep{0pt}%
    \put(0,0){\includegraphics[width=\unitlength,page=1]{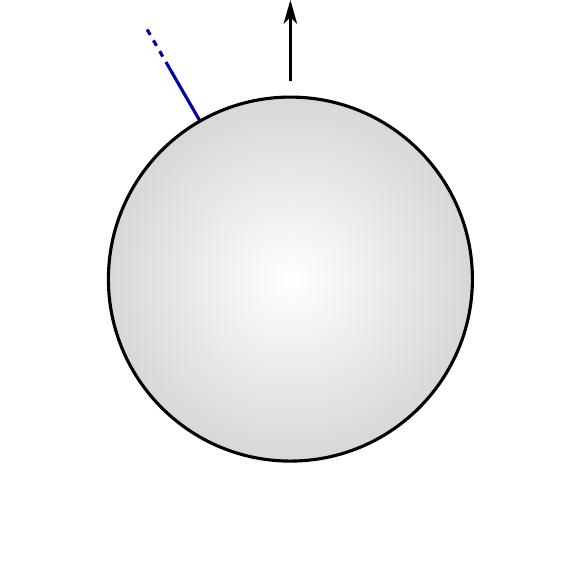}}%
    \put(0.52166575,0.70655597){\color[rgb]{0,0,0}\makebox(0,0)[lt]{\lineheight{1.25}\smash{\begin{tabular}[t]{l}inside\end{tabular}}}}%
    \put(0.5230626,0.893667){\color[rgb]{0,0,0}\makebox(0,0)[lt]{\lineheight{1.25}\smash{\begin{tabular}[t]{l}outside\end{tabular}}}}%
    \put(0,0){\includegraphics[width=\unitlength,page=2]{bubble1_with_strings.pdf}}%
  \end{picture}%
\endgroup%

		\caption{}
		\label{fig:bubble}
	\end{subfigure}%
	\hfill
	\begin{subfigure}[b]{0.45\textwidth}
		\centering
		\def\svgwidth{0.8\linewidth}
\begingroup%
  \makeatletter%
  \providecommand\color[2][]{%
    \errmessage{(Inkscape) Color is used for the text in Inkscape, but the package 'color.sty' is not loaded}%
    \renewcommand\color[2][]{}%
  }%
  \providecommand\transparent[1]{%
    \errmessage{(Inkscape) Transparency is used (non-zero) for the text in Inkscape, but the package 'transparent.sty' is not loaded}%
    \renewcommand\transparent[1]{}%
  }%
  \providecommand\rotatebox[2]{#2}%
  \newcommand*\fsize{\dimexpr\f@size pt\relax}%
  \newcommand*\lineheight[1]{\fontsize{\fsize}{#1\fsize}\selectfont}%
  \ifx\svgwidth\undefined%
    \setlength{\unitlength}{349.4998613bp}%
    \ifx\svgscale\undefined%
      \relax%
    \else%
      \setlength{\unitlength}{\unitlength * \real{\svgscale}}%
    \fi%
  \else%
    \setlength{\unitlength}{\svgwidth}%
  \fi%
  \global\let\svgwidth\undefined%
  \global\let\svgscale\undefined%
  \makeatother%
  \begin{picture}(1,0.65314418)%
    \lineheight{1}%
    \setlength\tabcolsep{0pt}%
    \put(0.22949824,0.17011173){\color[rgb]{0,0,0}\makebox(0,0)[t]{\lineheight{1.25}\smash{\begin{tabular}[t]{c}inside\end{tabular}}}}%
    \put(0.82071514,0.4993038){\color[rgb]{0,0,0}\makebox(0,0)[t]{\lineheight{1.25}\smash{\begin{tabular}[t]{c}outside\end{tabular}}}}%
    \put(0,0){\includegraphics[width=\unitlength,page=1]{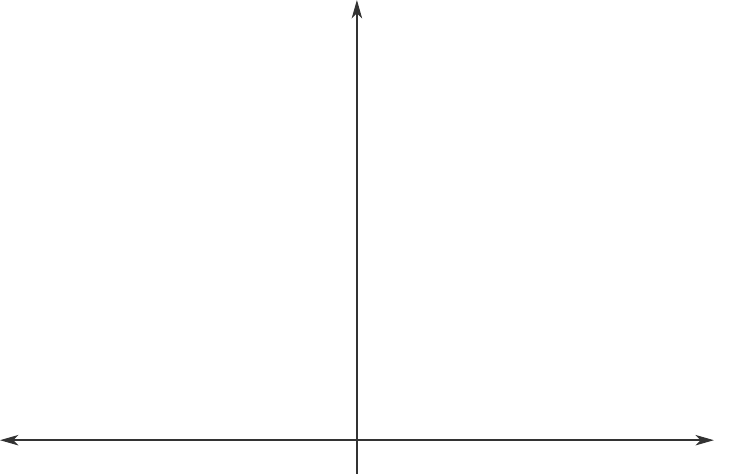}}%
    \put(0.97199402,0.00504627){\color[rgb]{0,0,0}\makebox(0,0)[t]{\lineheight{1.25}\smash{\begin{tabular}[t]{c}$z$\end{tabular}}}}%
    \put(0,0){\includegraphics[width=\unitlength,page=2]{figures/shell_warp.pdf}}%
    \put(0.67220218,0.17011173){\color[rgb]{0,0,0}\makebox(0,0)[lt]{\lineheight{1.25}\smash{\begin{tabular}[t]{l}Randall-Sundrum\end{tabular}}}}%
    \put(0.43117848,0.59861845){\color[rgb]{0,0,0}\makebox(0,0)[t]{\lineheight{1.25}\smash{\begin{tabular}[t]{c}$e^{A(z)}$\end{tabular}}}}%
  \end{picture}%
\endgroup%

		\caption{}
		\label{fig:scale_factor}
	\end{subfigure}
	\caption{(a) Radially stretched strings (in blue) attached to the expanding bubble correspond to massive particles (in red) in the four dimensional universe on the dark bubble. (b) The volume of radial slices decreases inward and increases outward, defining the inside and the outside of the bubble respectively.}
	\label{fig:scale_factor_and_bubble}
\end{figure*}

\section*{Life on an expanding bubble}

A common approach to deal with extra dimensions in string theory, is to first find a time independent compactification and then extract a four dimensional low energy effective field theory. With the help of this theory one can then search for time dependent solutions. In this way, particle physics precedes cosmology. As we have seen, the swampland program suggests that this approach might fail. 

We propose that cosmology plays a key role already at a fundamental level, and that the naive picture of our universe as an expanding balloon in an extra dimension is literally true. Ironically, the vacuum instabilities making it difficult to find dS vacua within the usual paradigm are exactly what we need to make our model work. 

The idea put forth in \cite{Banerjee:2018qey}, is that our universe is riding on a 1+3 dimensional bubble in a 1+4 dimensional AdS spacetime. As observed in, e.g. \cite{Brown:1988kg}, an AdS$_5$ vacuum 
can decay via the nucleation of bubbles of true vacuum, corresponding to an AdS$_5$ with an even more negative vacuum energy. The Israel junction conditions imply

\be
\label{eq:junct}
\sigma=\frac{3}{8πG_5}\left(\sqrt{k^2_{-}+\frac{1+\dot{a}^2}{a^2}}-\sqrt{k_{+}^2+\frac{1+\dot{a}^2}{a^2}}\right)\,,
\ee
where $\sigma$ is the tension of the brane supporting the shell, while the  negative cosmological constants on the inside and the outside are $\Lambda_{\mp}$ with $\Lambda_- < \Lambda _+<0$, where $\Lambda_\pm=-6k^2_\pm$. Their magnitudes are assumed to be characterized by high energy physics. The induced metric on the shell has a FLRW form with scale factor $a(\tau)$,
\be
ds²_{\mathrm{shell}}=-dτ²+a(τ)²dΩ²₃ \, .
\ee
For large $k_\pm$ we find
\be
\frac{\dot{a}²}{a²} \approx-\frac{1}{a²}+\frac{8πG₄}{3}Λ₄\,,
\ee
which is just the Friedmann equation (with the four dimensional Newton's constant obtained from the five dimensional one through $G_4=\frac{2k_- k_+ }{k_- -k_+} G_5$) in the presence of a positive cosmological constant $\Lambda_4$ given by 
\be
\Lambda_4 = \frac{3\left(k_{-}-k₊\right)}{8πG_5 }- \sigma.  
\ee
Gravity on the shell is described by four dimensional Einstein's equations plus high energy corrections \cite{Banerjee:2019fzz}. Further aspects were explored in \cite{Banerjee:2020wix,Banerjee:2020wov,Banerjee:2021qei}.


One can generate radiation on the four dimensional world through a combination of radiation on the bubble and a nontrivial metric inside and outside of the shell. In the presence of an AdS-Schwarzschild metric with ADM mass, $M_{\pm}$, the Friedmann equation becomes
\be
\frac{\dot{a}²}{a²} \approx -\frac{1}{a²}+\frac{8πG₄}{3}\left[Λ₄+\frac{1}{2π²a^4}\left(\frac{M₊}{k₊}-\frac{M_-}{k_-}\right)\right].
\ee
We see, for instance, how $M_+>0$ on the outside, but $M_-=0$ on the inside leads to a positive density of radiation. 

\section*{Spaghetti in the hemlock}

In our construction, bulk strings are the most essential ingredients, which, 
in conjunction with certain instabilities and no-go theorems in supergravity,
provide a delicious recipe for a ride out of the swamp. Let us summarize the main achievements of our model.

\begin{itemize}
\item As was shown in \cite{Banerjee:2020wix}, to get a massive particle, one needs a string that pulls upwards from the brane as in \fref{fig:bubble}.
With a homogeneous distribution of many such strings, one reproduces the Friedmann equations in the presence of dust. What happens is that the dark bubble eats the strings as it expands. The energy from the strings is used so that the effective energy density on the bubble, and hence $H^2$, decays only as $1/a^3$ rather than $1/a^4$ as in the case of radiation. 
\be
\begin{split}	
\label{Friedmann-strings}
 H^2 \equiv &\frac{\dot{a}^2}{a^2} \approx -\frac{1}{a^2} + \frac{8\pi G_4}{3} \left[\Lambda_4 + \phantom{\frac{1}{π}} \right.\\ 
 &\left. \frac{1}{2\pi^2a^4} \left(\frac{M_+}{k_+}-\frac{M_-}{k_-}\right)
+ \frac{3}{8\pi a^3} \frac{\tau}{k_+}  \right].
\end{split}
\ee
In our model, four dimensional physics is not localized to the shell; rather, it is described by the full five dimensional bulk in which the strings stretch. It is evident from \eqref{Friedmann-strings} that the effective mass of a particle is given by $τ L_\text{AdS}$, where $\tau$ is the tension of the string, and $L_\text{AdS}=1/k_+$. Hence the mass is independent of the length of the string  \cite{Banerjee:2019fzz, Banerjee:2020wix}.
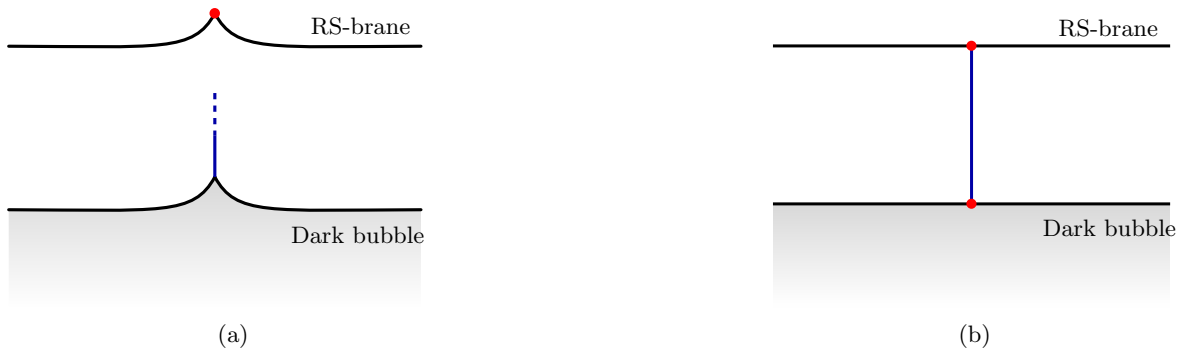
\begin{figure*}
	\centering
	\begin{subfigure}[b]{0.45\textwidth}
		\centering
		\def\svgwidth{0.75\linewidth}
\begingroup%
  \makeatletter%
  \providecommand\color[2][]{%
    \errmessage{(Inkscape) Color is used for the text in Inkscape, but the package 'color.sty' is not loaded}%
    \renewcommand\color[2][]{}%
  }%
  \providecommand\transparent[1]{%
    \errmessage{(Inkscape) Transparency is used (non-zero) for the text in Inkscape, but the package 'transparent.sty' is not loaded}%
    \renewcommand\transparent[1]{}%
  }%
  \providecommand\rotatebox[2]{#2}%
  \newcommand*\fsize{\dimexpr\f@size pt\relax}%
  \newcommand*\lineheight[1]{\fontsize{\fsize}{#1\fsize}\selectfont}%
  \ifx\svgwidth\undefined%
    \setlength{\unitlength}{161.32818407bp}%
    \ifx\svgscale\undefined%
      \relax%
    \else%
      \setlength{\unitlength}{\unitlength * \real{\svgscale}}%
    \fi%
  \else%
    \setlength{\unitlength}{\svgwidth}%
  \fi%
  \global\let\svgwidth\undefined%
  \global\let\svgscale\undefined%
  \makeatother%
  \begin{picture}(1,0.65779316)%
    \lineheight{1}%
    \setlength\tabcolsep{0pt}%
    \put(0,0){\includegraphics[width=\unitlength,page=1]{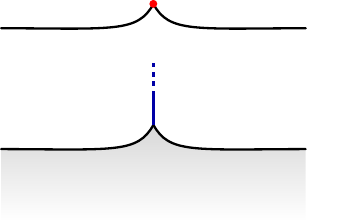}}%
    \put(0.66731246,0.60119604){\color[rgb]{0,0,0}\makebox(0,0)[lt]{\lineheight{1.25}\smash{\begin{tabular}[t]{l}RS-brane\end{tabular}}}}%
    \put(0.62385001,0.14156876){\color[rgb]{0,0,0}\makebox(0,0)[lt]{\lineheight{1.25}\smash{\begin{tabular}[t]{l}Dark bubble\end{tabular}}}}%
  \end{picture}%
\endgroup%

		\caption{}
		\label{fig:sandwicha}
	\end{subfigure}%
	\hfill
	\begin{subfigure}[b]{0.45\textwidth}
		\centering
		\def\svgwidth{0.67\linewidth}
\begingroup%
  \makeatletter%
  \providecommand\color[2][]{%
    \errmessage{(Inkscape) Color is used for the text in Inkscape, but the package 'color.sty' is not loaded}%
    \renewcommand\color[2][]{}%
  }%
  \providecommand\transparent[1]{%
    \errmessage{(Inkscape) Transparency is used (non-zero) for the text in Inkscape, but the package 'transparent.sty' is not loaded}%
    \renewcommand\transparent[1]{}%
  }%
  \providecommand\rotatebox[2]{#2}%
  \newcommand*\fsize{\dimexpr\f@size pt\relax}%
  \newcommand*\lineheight[1]{\fontsize{\fsize}{#1\fsize}\selectfont}%
  \ifx\svgwidth\undefined%
    \setlength{\unitlength}{216.41903504bp}%
    \ifx\svgscale\undefined%
      \relax%
    \else%
      \setlength{\unitlength}{\unitlength * \real{\svgscale}}%
    \fi%
  \else%
    \setlength{\unitlength}{\svgwidth}%
  \fi%
  \global\let\svgwidth\undefined%
  \global\let\svgscale\undefined%
  \makeatother%
  \begin{picture}(1,0.70903846)%
    \lineheight{1}%
    \setlength\tabcolsep{0pt}%
    \put(0,0){\includegraphics[width=\unitlength,page=1]{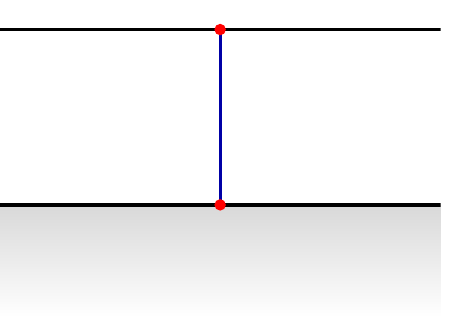}}%
    \put(0.70240025,0.66691099){\color[rgb]{0,0,0}\makebox(0,0)[lt]{\lineheight{1.25}\smash{\begin{tabular}[t]{l}RS-brane\end{tabular}}}}%
    \put(0.66352169,0.17495377){\color[rgb]{0,0,0}\makebox(0,0)[lt]{\lineheight{1.25}\smash{\begin{tabular}[t]{l}Dark bubble\end{tabular}}}}%
  \end{picture}%
\endgroup%

		\caption{}
		\label{fig:sandwichc}
	\end{subfigure}
	\caption{(a) A string ending on the dark bubble and a point mass placed on a RS-brane, both bend away from the true vacuum (in gray). (b) When the stretched string ends on the RS-brane, a change of gauge gives a sandwich universe.}
	\label{fig:sandwich}
\end{figure*}
\item The strings are essential for extracting a consistent effective theory of gravity on the brane. The value of $G_4$ is finite and independent of the possibly infinite volume of the extra dimensions. Previous braneworld models following Randall-Sundrum (RS) \cite{Randall:1999ee,Randall:1999vf}, primarily dealt with warped extra dimension corresponding to the gluing of two insides of a bubble (see \fref{fig:scale_factor}). 
In the absence of the stringy sources that are unique to our model, there is no other simple way to render $G_4$  finite \cite{Banerjee:2020wov}.

\item The way the low energy dynamics on the brane is imprinted by large extra dimensions, avoids several no-go theorems \cite{Maldacena:2000mw} that strictly prevent dS embedding with compact extra dimensions and positive tension sources.

\item A false vacuum in the form of a non-supersymmetric AdS can decay \cite{Ooguri:2016pdq, Freivogel:2016qwc} via the nucleation of spherical bubbles of \emph{true} vacuum \cite{Coleman:1980aw}. This provides a natural realization of the scenario described thus far. 

Therefore, the instability that threatens to destroy any hope for a non-supersymmetric vacuum in string theory, can rescue string theory out of the cosmological swampland and be the key to realizing a dS universe.
\end{itemize}

\section*{A sandwich universe?}

To provide a holographic interpretation of our model, we need to insert a cutoff brane at large $z$, corresponding to the UV in a holographic interpretation. We should do this so that the Einstein equations are satisfied on the branes and in the bulk. An interesting way to achieve this is through a RS-brane with $\mathbb{Z}_2$ symmetry across it, and consequently two insides. With the tension tuned to $\sigma = 2 k_+$ the brane is flat. 

Let us consider strings stretching between the dark bubble and the cutoff RS-brane. A stretched string pulls on the dark bubble as shown in black in \fref{fig:sandwicha}, and induces Schwarzschild geometry on it. On the other hand, a point mass placed on the RS-brane causes it to bend away from its inside, again inducing Schwarzschild. We can then make a change of gauge so that both the branes are flat and the metric in the bulk is that of a black string \cite{Chamblin:1999by}
\be
\begin{split}
	\D s^2 = 
	&\frac{L^2}{z^2} \left[- \left(1-\frac{2 m}{r}\right) \D t^2 + \left(1-\frac{2 m}{r}\right)^{-1} \D r^2 \right. \\ 
	 & \left. \phantom{\frac{2 m}{r}}+ r^2 \D \Omega_2^2 + \D z^2 \right].
\end{split}
\ee
The change of gauge is singular and induces a point mass on the dark bubble located at a constant $z$, and a string pulling on the RS-brane. The final result is shown in \fref{fig:sandwich} with strings and masses pulling in a way that leads to flat branes with induced Schwarzschild metrics.

\begin{figure*}[t]
	\begin{subfigure}[b]{0.45\textwidth}
		\centering
		\includegraphics[width=0.7\textwidth]{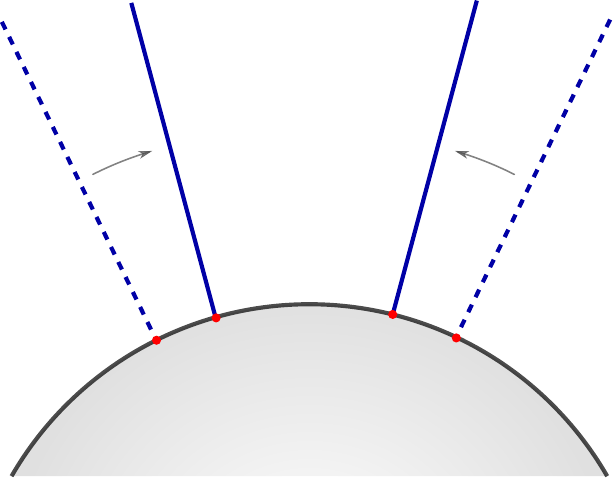}
		\caption{}
		\label{fig:radial_strings1}
	\end{subfigure}
	\hfill
	\begin{subfigure}[b]{0.45\textwidth}
		\centering
		\includegraphics[width=0.7\textwidth]{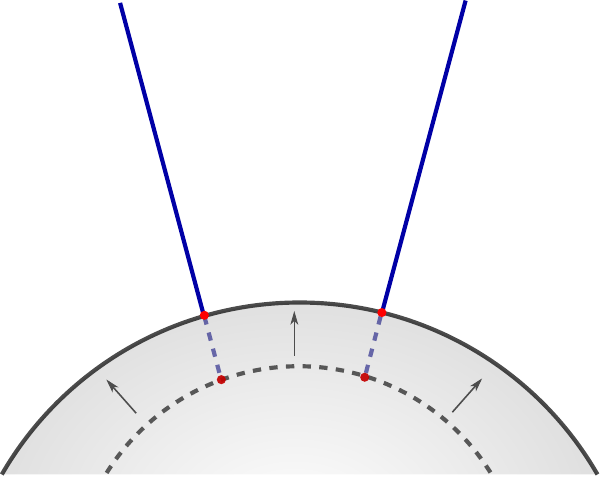}
		\caption{}
		\label{fig:radial_strings2}
	\end{subfigure}
	\caption{(a) Strings stretching radially outwards from the shell gravitate towards each other, while remaining radial. (b) In the limit where the gravitational interaction between them can be ignored, the strings get ``eaten up'' by the expanding shell.}
   	\label{fig:radial_strings}
\end{figure*}
In the limit where the motion of the dark bubble and the expansion of the universe can be ignored, the strings move in the bulk geometry in such a way they remain radially stretched. In \cite{Banerjee:2020wov} it was shown that this projects to geodesic motion on the dark bubble. If we ignore the force between the strings, and focus on the expansion, we see how the proper distance between the endpoints of the strings on the dark bubble grows (\fref{fig:radial_strings}). The end points on the RS-brane remain constant and correspond to comoving coordinates. For a gravitationally bound system, the proper size remains constant even as the universe expands. This means that the dark bubble pulls the strings together as it expands. This, in turn, means that the size of the bound system {\it decreases} on the RS-brane. The RS-brane gives a different, but equivalent picture of the time evolution where the universe does not expand. Instead, all physical scales become smaller. 
One can think of two different ways to realize this model in AdS/CFT. 
\begin{enumerate}
\item One way is to take the cutoff RS-brane very close to the boundary of AdS. In that case the stress tensor induced on the brane will have contributions coming from the non-normalizable  bulk strings which imprint local massive deformations on the brane, as well as from the holographic stress-energy tensor \cite{deHaro:2000wj}.

\item In a realistic scenario, nucleation of other bubbles in our AdS, might actually compel us to settle for a finite cutoff. In this case we need to consider, instead, the holographic correspondence between the cutoff AdS and the $TT$-deformed CFTs \cite{McGough:2016lol, Hartman:2018tkw, Taylor:2018xcy, Guica:2019nzm}. 
\end{enumerate}

\section*{Conclusion}

The dark bubble is a new way of realizing dS in string theory, which makes use of key features of string theory such as extra dimensions, AdS spacetimes and instabilites. The framework suggests that dark energy, and possibly also other dark components of matter, are features of higher dimensional physics. 

\small
\bibliography{references}
\bibliographystyle{utphysmodb}
\end{document}